\documentclass[twocolumn,showpacs,fleqn,nobibnotes]{revtex4}

\usepackage{amsmath}
\usepackage{graphicx}
\usepackage{float}
\usepackage{subfigure}

\def\lsim{\raise0.3ex\hbox{$<$\kern-0.75em\raise-1.1ex\hbox{$\sim$}}}
\def\gsim{\raise0.3ex\hbox{$>$\kern-0.75em\raise-1.1ex\hbox{$\sim$}}}

\newcommand{\rr}{\mbox{\boldmath $r$}}

\newcommand{\rb}{\mbox{\boldmath $b$}}
\newcommand{\rd}{\mbox{\boldmath $\Delta$}}

\begin{document}

\title{Vector Meson Production in Coherent Hadronic Interactions: An update on predictions for RHIC and LHC}
\author{V.P. Gon\c{c}alves
$^{a}$
and M.V.T. Machado $^{b}$ }

\affiliation{$^a$ Instituto de F\'{\i}sica e Matem\'atica, Universidade Federal de
Pelotas\\
Caixa Postal 354, CEP 96010-900, Pelotas, RS, Brazil.\\
$^b$ Instituto de F\'{\i}sica, Universidade Federal do Rio Grande do Sul\\ Caixa Postal 15051, CEP 91501-970, Porto Alegre, RS, Brazil.}

\begin{abstract}
In this letter we update our predictions for  the photoproduction of vector mesons in coherent $pp$ and $AA$  collisions at RHIC and LHC energies using the color dipole approach and  the Color Glass Condensate (CGC) formalism. In particular, we present our predictions for the first run of the LHC at half energy and for  the rapidity dependence of the ratio between the $J/\Psi$ and $\rho$ cross sections at RHIC energies. 

\end{abstract}

\maketitle


The Large Hadron Collider (LHC) at CERN  started nineteen months ago. During this period a large amount of data have been collected considering $pp$ collisions at $\sqrt{s}$ = 0.9, 2.36 and 7 TeV as well as $PbPb$ collisions at $\sqrt{s}$ = 2.76 TeV.  Currently, there is a great
expectation that LHC will discover the Higgs boson and whatever new physics beyond the Standard Model that may accompany it, such as
supersymmetry or extra dimensions \cite{lhc}. However, we should remember that LHC opens a new kinematical regime at high energy, where several questions related to the description of the high-energy regime of the Quantum
Chromodynamics (QCD) remain without  satisfactory answers (For recent reviews see Ref. \cite{hdqcd}).
In recent years we have proposed the analysis of coherent collisions in hadronic interactions as an alternative way to study the QCD dynamics at high energies
\cite{vicmag_upcs,vicmag_hq,vicmag_mesons_per,vicmag_prd,vicmag_pA,vicmag_difper,vicmag_quarkonium,vicmag_rho,vicmag_ane}. The basic idea in coherent  hadronic collisions is that
the total cross section for a given process can be factorized in
terms of the equivalent flux of photons into the hadron projectile and
the photon-photon or photon-target production cross section.
The main advantage of using colliding hadrons and nuclear beams for
studying photon induced interactions is the high equivalent photon
energies and luminosities that can be achieved at existing and
future accelerators (For recent reviews see Ref. \cite{upcs}).
 Consequently, studies of $\gamma p(A)$ interactions
at the LHC could provide valuable information on the QCD dynamics at
high energies.

Our goal in this letter is to furnish an update of our predictions for the vector meson production in coherent $pp$ and $AA$ collisions which can be used in the current and future  analysis of the  experimental data from RHIC and LHC. In particular, we will update the Refs. \cite{vicmag_quarkonium,vicmag_rho}, where we have studied the $J/\Psi,\,\Upsilon$ and $\rho$ production at RHIC and for the full energies of the LHC. Moreover, for the first time, we will present our predictions for the rapidity dependence of the ratio between the $J/\Psi$ and $\rho$ cross sections at RHIC energies, which is currently under analysis by the STAR collaboration. Before presenting our results, in what follows we will introduce a very brief explanation about the models used in our calculations and refer to the Refs. \cite{vicmag_quarkonium,vicmag_rho} for more complete details. 

Lets consider the hadron-hadron interaction at large impact parameter ($b > 2R_{h}$, where $R_h$ is the hadron radius) and at ultra relativistic energies. In this regime we expect the electromagnetic interaction to be dominant.
In  heavy ion colliders, the heavy nuclei give rise to strong electromagnetic fields due to the coherent action of all protons in the nucleus, which can interact with each other. In a similar way, it also occurs when considering ultra relativistic  protons in $pp(\bar{p})$ colliders.
The cross section for the photoproduction of a vector meson $V$ in a coherent  hadron-hadron collision is  given by,
\begin{eqnarray}
\sigma (h h \rightarrow V h) =2 \int \limits_{\omega_{min}}^{\infty} d\omega \int dt \,\frac{dN_{\gamma}(\omega)}{d\omega}\,\frac{d\sigma}{dt} \left(W_{\gamma h},t\right)\,,
\label{sigAA}
\end{eqnarray}
where$\frac{dN_{\gamma}\,(\omega)}{d\omega}$ is the equivalent photon flux, $\frac{d\sigma}{dt}$ is the differential cross section for the process    $(\gamma h \rightarrow V h)$, $\omega_{min}=M_{V}^2/4\gamma_L m_p$, $W_{\gamma h}^2=2\,\omega\sqrt{S_{\mathrm{NN}}}$  and
$\sqrt{S_{\mathrm{NN}}}$ is  the c.m.s energy of the
hadron-hadron system.
Considering the requirement that  photoproduction
is not accompanied by hadronic interaction (ultra-peripheral
collision) an analytic approximation for the equivalent photon flux of a nuclei can be calculated, which is given by \cite{upcs}
\begin{eqnarray}
\frac{dN_{\gamma}\,(\omega)}{d\omega}= \frac{2\,Z^2\alpha_{em}}{\pi\,\omega}\, \left[\bar{\eta}\,K_0\,(\bar{\eta})\, K_1\,(\bar{\eta})+ \frac{\bar{\eta}^2}{2}\,{\cal{U}}(\bar{\eta}) \right]\,
\label{fluxint}
\end{eqnarray}
where
 $\omega$ is the photon energy,  $\gamma_L$ is the Lorentz boost  of a single beam and $\eta
= \omega b/\gamma_L$; $K_0(\eta)$ and  $K_1(\eta)$ are the
modified Bessel functions.
Moreover, $\bar{\eta}=\omega\,(2.R_{h})/\gamma_L$ and  ${\cal{U}}(\bar{\eta}) = K_1^2\,(\bar{\eta})-  K_0^2\,(\bar{\eta})$.
 On the other hand, for   proton-proton interactions, we assume that the  photon spectrum is given by  \cite{Dress},
\begin{eqnarray}
\frac{dN_{\gamma}(\omega)}{d\omega} =  \frac{\alpha_{\mathrm{em}}}{2 \pi\, \omega} \left[ 1 + \left(1 -
\frac{2\,\omega}{\sqrt{S_{NN}}}\right)^2 \right] \nonumber \\
\times \left( \ln{\Omega} - \frac{11}{6} + \frac{3}{\Omega}  - \frac{3}{2 \,\Omega^2} + \frac{1}{3 \,\Omega^3} \right) \,,
\label{eq:photon_spectrum}
\end{eqnarray}
with the notation $\Omega = 1 + [\,(0.71 \,\mathrm{GeV}^2)/Q_{\mathrm{min}}^2\,]$ and $Q_{\mathrm{min}}^2= \omega^2/[\,\gamma_L^2 \,(1-2\,\omega /\sqrt{S_{NN}})\,] \approx (\omega/
\gamma_L)^2$.
The factor two in Eq. (\ref{sigAA}) takes into account the fact that the hadron can act as both target and photon emitter. The experimental separation for such events is relatively easy, as photon emission is coherent over the  hadron and the photon is colorless we expect the events to be characterized by intact recoiled hadron (tagged hadron) and a two rapidity gap pattern (For a detailed discussion see \cite{upcs}).

We describe the vector meson production in the color dipole frame, in which most of the energy is
carried by the hadron, while the  photon  has
just enough energy to dissociate into a quark-antiquark pair
before the scattering. In this representation the probing
projectile fluctuates into a
quark-antiquark pair (a dipole) with transverse separation
$\rr$ long after the interaction, which then
scatters off the hadron \cite{nik}.
In the dipole picture the   amplitude for production of a  vector meson $V$ is given by (See e.g. Refs. \cite{nik,vicmag_mesons,KMW})
\begin{eqnarray}
\, {\cal A}_{T,L}^{\gamma^*h \rightarrow V h}\, (x,Q^2,\Delta)  = 
\int dz\, d^2\rr \,(\Psi^{V*}\Psi)_{T,L}\,{\cal{A}}_{q\bar{q}}(x,\rr,\Delta) \, ,
\label{sigmatot}
\end{eqnarray}
where $(\Psi^{V*}\Psi)_{T,L}$ denotes the overlap of the photon and vector meson wave functions. The variable  $z$ $(1-z)$ is the
longitudinal momentum fractions of the quark (antiquark),  $\Delta$ denotes the transverse momentum lost by the outgoing hadron ($t = - \Delta^2$) and $x$ is the Bjorken variable.  Moreover, ${\cal{A}}_{q\bar{q}}$ is the elementary elastic amplitude for the scattering of a dipole of size $\rr$ on the target. It is directly related to scattering amplitude ${\cal{N}} (x,\rr,\rb)$ and consequently to the QCD dynamics (see below).  One has that \cite{KMW}
\begin{eqnarray}
{\cal{A}}_{q\bar{q}} (x,\rr,\Delta) & = & i \int d^2 \rb \, e^{-i \rb.\rd}\, 2 {\cal{N}}(x,\rr,\rb) \,\,,
\end{eqnarray}
where $\rb$ is the transverse distance from the center of the target to one of the $q \bar{q}$ pair of the dipole.  Consequently, one can express the amplitude for the exclusive production of a vector meson as follows
\begin{eqnarray}
 {\cal A}_{T,L}^{\gamma^*h \rightarrow V h}(x,Q^2,\Delta) & = & i
\int dz \, d^2\rr \, d^2\rb  e^{-i[\rb-(1-z)\rr].\rd} \nonumber \\
 &\times & (\Psi_{V}^* \Psi)_T \,2 {\cal{N}}(x,\rr,\rb)
\label{sigmatot2}
\end{eqnarray}
where  the factor $[i(1-z)\rr].\rd$ in the exponential  arises when one takes into account non-forward corrections to the wave functions \cite{non}.
Finally, the differential cross section  for  exclusive vector meson production is given by
\begin{eqnarray}
\frac{d\sigma_{T,L}}{dt} (\gamma^* h \rightarrow V h) = \frac{1}{16\pi} |{\cal{A}}_{T,L}^{\gamma^*p \rightarrow V h}(x,Q^2,\Delta)|^2\,(1 + \beta^2)\,,
\label{totalcs}
\end{eqnarray}
where $\beta$ is the ratio of real to imaginary parts of the scattering
amplitude. For the case of heavy mesons, skewness corrections are quite important and they are also taken  into account. (For details, see Refs. \cite{vicmag_mesons,KMW}).

\begin{figure}[t]
\includegraphics[scale=0.15]{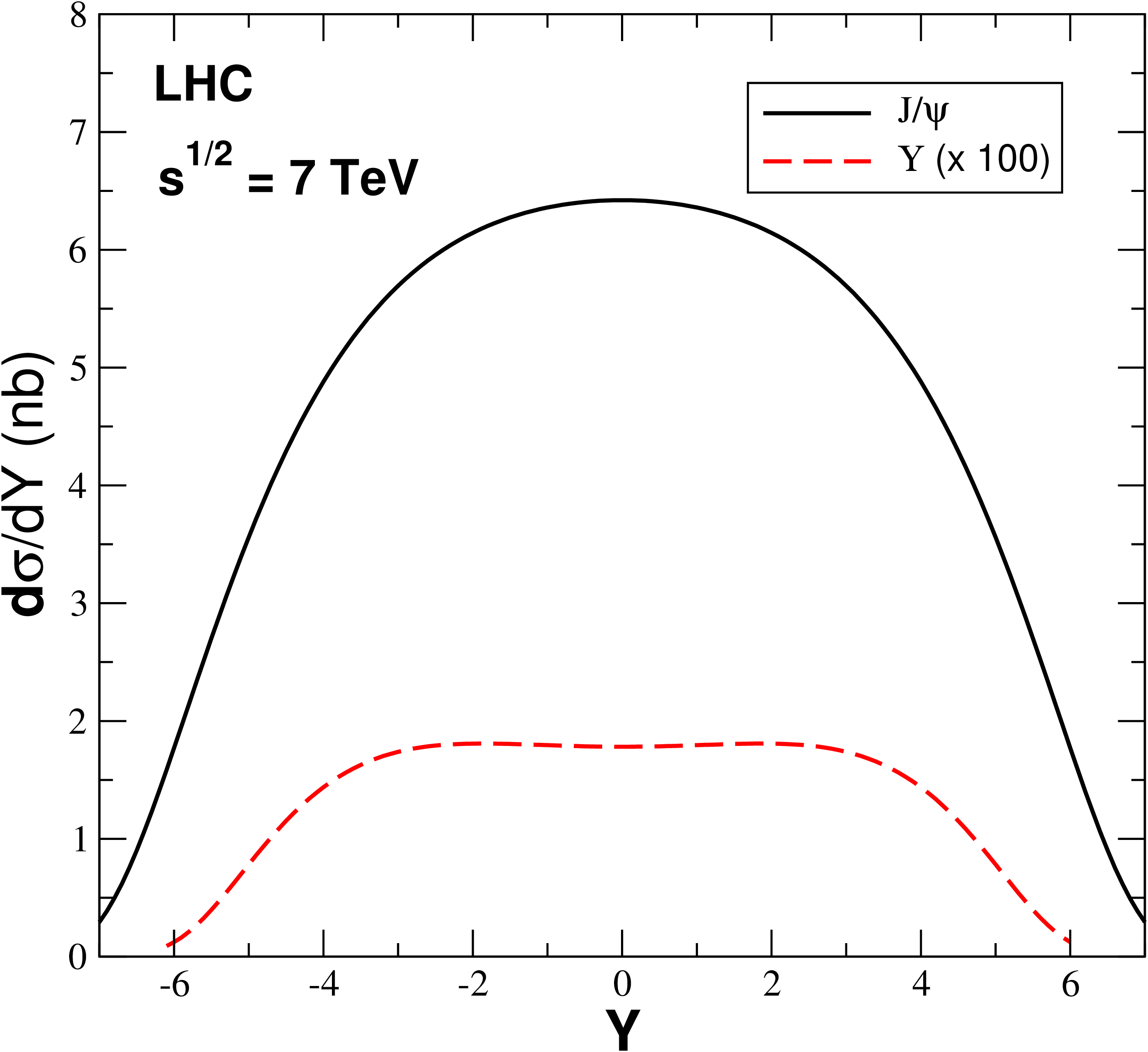}
 \caption{(Color online) Predictions for the rapidity distribution of $J/\Psi$ and $\Upsilon$ photoproduction in $pp$ collisions at LHC ($\sqrt{s} = 7$ TeV).}
\label{fig:1}
\end{figure}

The scattering amplitude ${\cal{N}}(x,\rr,\rb)$ contains all
information about the target and the strong interaction physics.
In the Color Glass Condensate (CGC)  formalism \cite{CGC,BAL,WEIGERT}, it  encodes all the
information about the
non-linear and quantum effects in the hadron wave function.
It can be obtained by solving an appropriate evolution
equation in the rapidity $y\equiv \ln (1/x)$. As in Refs. \cite{vicmag_quarkonium,vicmag_rho} we will assume that in the case of photon-nucleus interactions   the scattering amplitude  is given by \cite{armesto}
\begin{eqnarray}
{\cal{N}}(x,\rr,\rb) =  \left\{\, 1- \exp \left[-\frac{1}{2} A T_A(b)\, \sigma_{dip}(x,r)  \right] \right\}\,, \nonumber
\end{eqnarray}
where $T_A(b)$ is the nuclear profile function (obtained from a 3-parameter Fermi distribution for the nuclear
density) and  the dipole-nucleon cross section is taken from the Iancu, Itakura and Munier (IIM) model \cite{IIM}.

In the case of photon-proton interactions we will use the
non-forward saturation model of Ref. \cite{MPS} (hereafter MPS model), which captures the main features of the dependence on energy,  virtual photon virtuality and momentum transfer $t$ and describes quite well the HERA data  \cite{vicmagane_exclusive}.  In the MPS model, the elementary elastic amplitude for dipole interaction is given by,
\begin{eqnarray}
\label{sigdipt}
\mathcal{A}_{q\bar q}(x,r,\Delta)= \sigma_0\,e^{-B|t|} {\cal{N}} \left(rQ_{\mathrm{sat}}(x,|t|),x\right),
\end{eqnarray}
with the asymptotic behaviors $Q_{\mathrm{sat}}^2(x,\Delta)\sim
\max(Q_0^2,\Delta^2)\,\exp[-\lambda \ln(x)]$. Specifically, the $t$ dependence of the saturation scale is parametrised as
\begin{eqnarray}
\label{qsatt}
Q_{\mathrm{sat}}^2\,(x,|t|)=Q_0^2(1+c|t|)\:\left(\frac{1}{x}\right)^{\lambda}\,, \end{eqnarray}
in order to interpolate smoothly between the small and intermediate transfer
regions. For the parameter $B$ we use the value $B=3.754$ GeV$^{-2}$ \cite{MPS}. Finally, the scaling function ${\cal{N}}$ is obtained from the forward saturation model \cite{IIM}.

\begin{figure*}[t]
\includegraphics[scale=0.15]{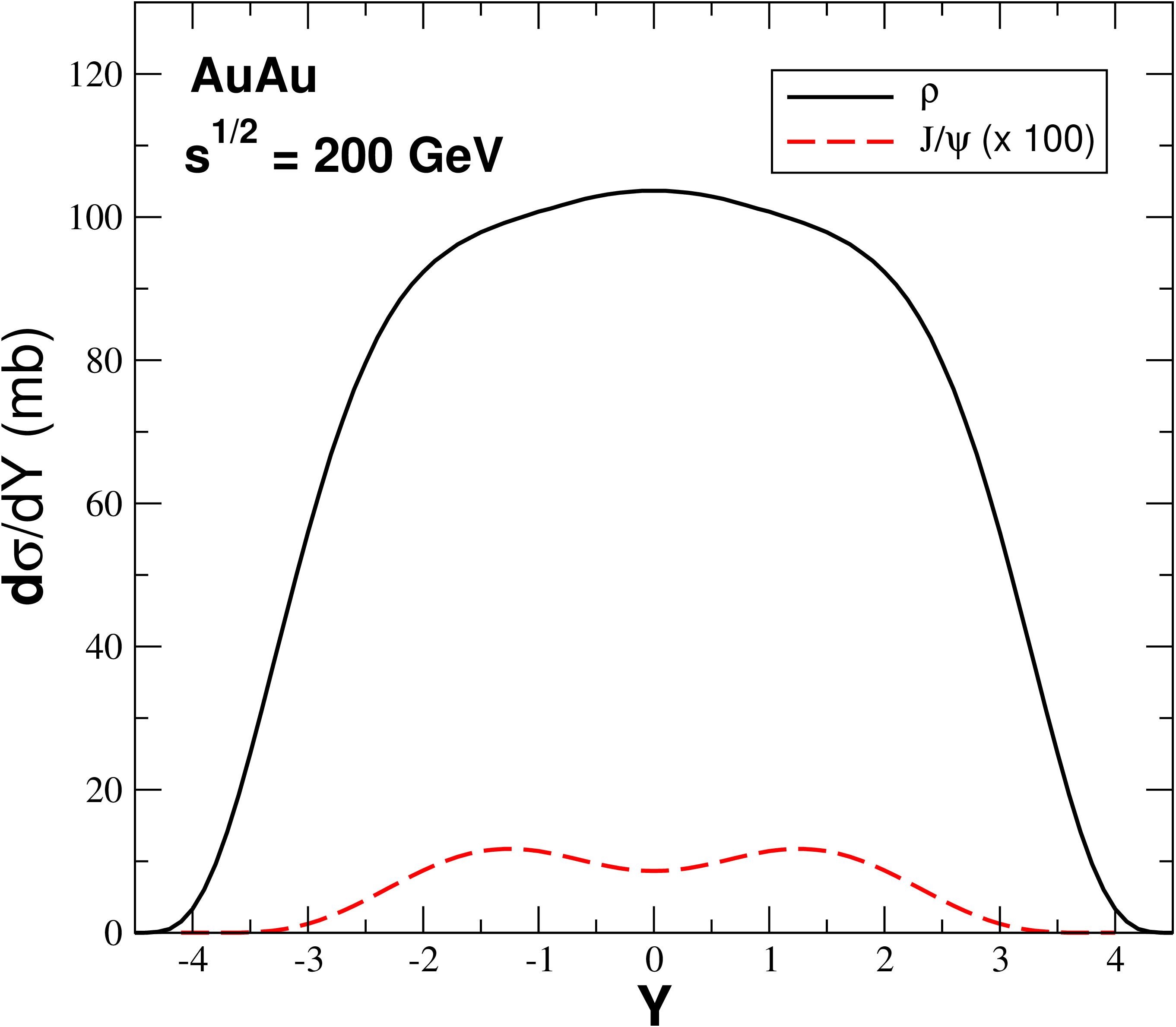} 
\hspace{0.4cm}
\includegraphics[scale=0.15]{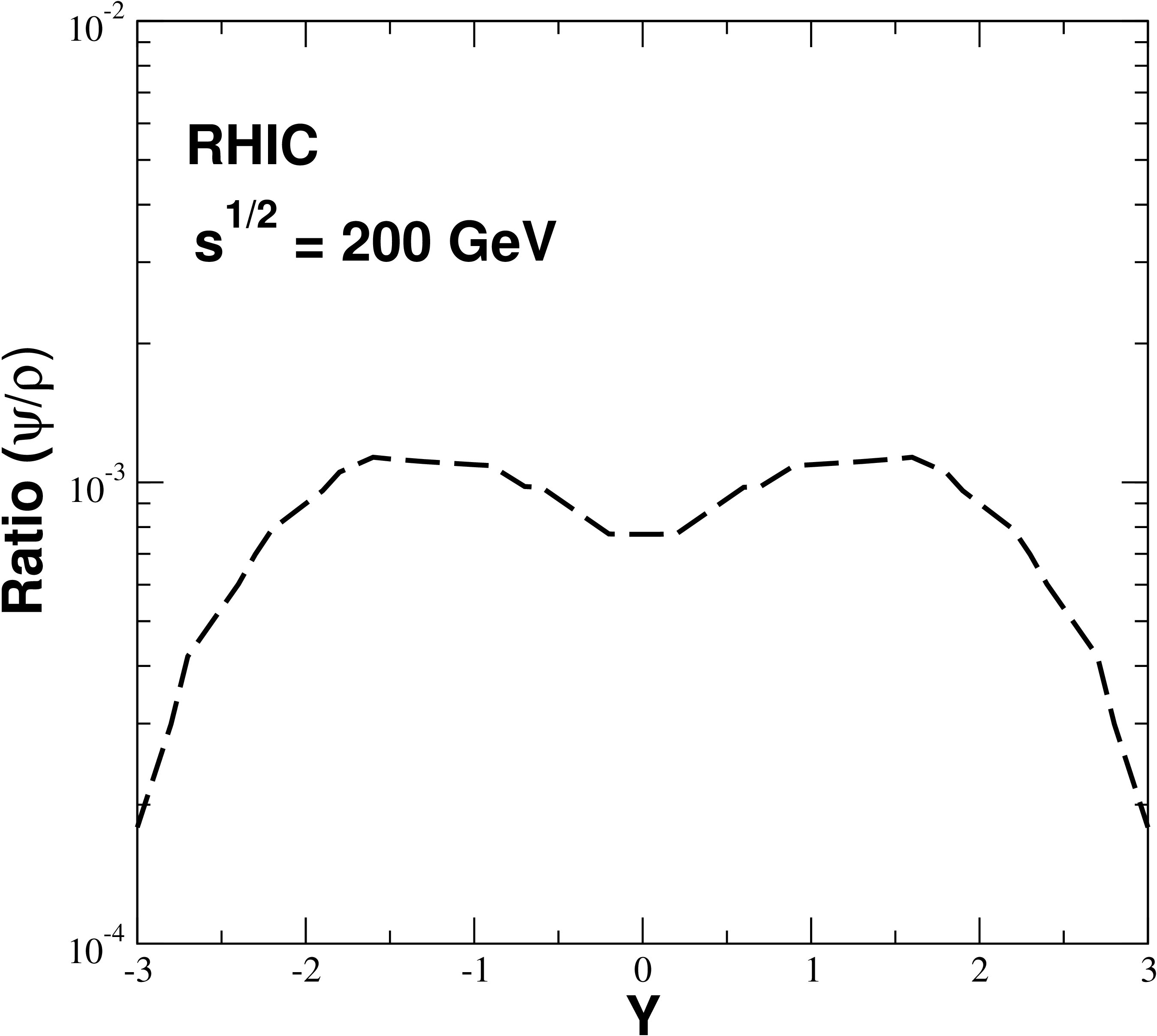}
 \caption{(Color online) Predictions for the rapidity distribution of $\rho$ and $J/\Psi$ photoproduction in $AuAu$ collisions at RHIC (left panel) and for the dependence on the   rapidity for the ratio between the $J/\Psi$ and $\rho$ cross sections (right panel).}
\label{fig:2}
\vspace{1cm}
\end{figure*}

Lets calculate the rapidity distribution and total cross section for the $J/\Psi$ and $\Upsilon$ photoproduction in  coherent $pp$ collisions.
The distribution on rapidity $Y$ of the produced final state can be directly computed from Eq. (\ref{sigAA}), by using its  relation with the photon energy $\omega$, i.e. $Y\propto \ln \, (2 \omega/M_{V})$.  Explicitly, the rapidity distribution is written down as,
\begin{eqnarray}
\frac{d\sigma \,\left[h + h \rightarrow   h \otimes V \otimes h \right]}{dY} = \omega \frac{dN_{\gamma} (\omega )}{d\omega }\,\sigma_{\gamma h \rightarrow V\, h}\left(\omega \right)\,
\label{dsigdy}
\end{eqnarray}
where $\otimes$ represents the presence of a rapidity gap. Consequently, given the photon flux, the rapidity distribution is thus a direct measure of the photoproduction cross section for a given energy.

\begin{figure}
\includegraphics[scale=0.15]{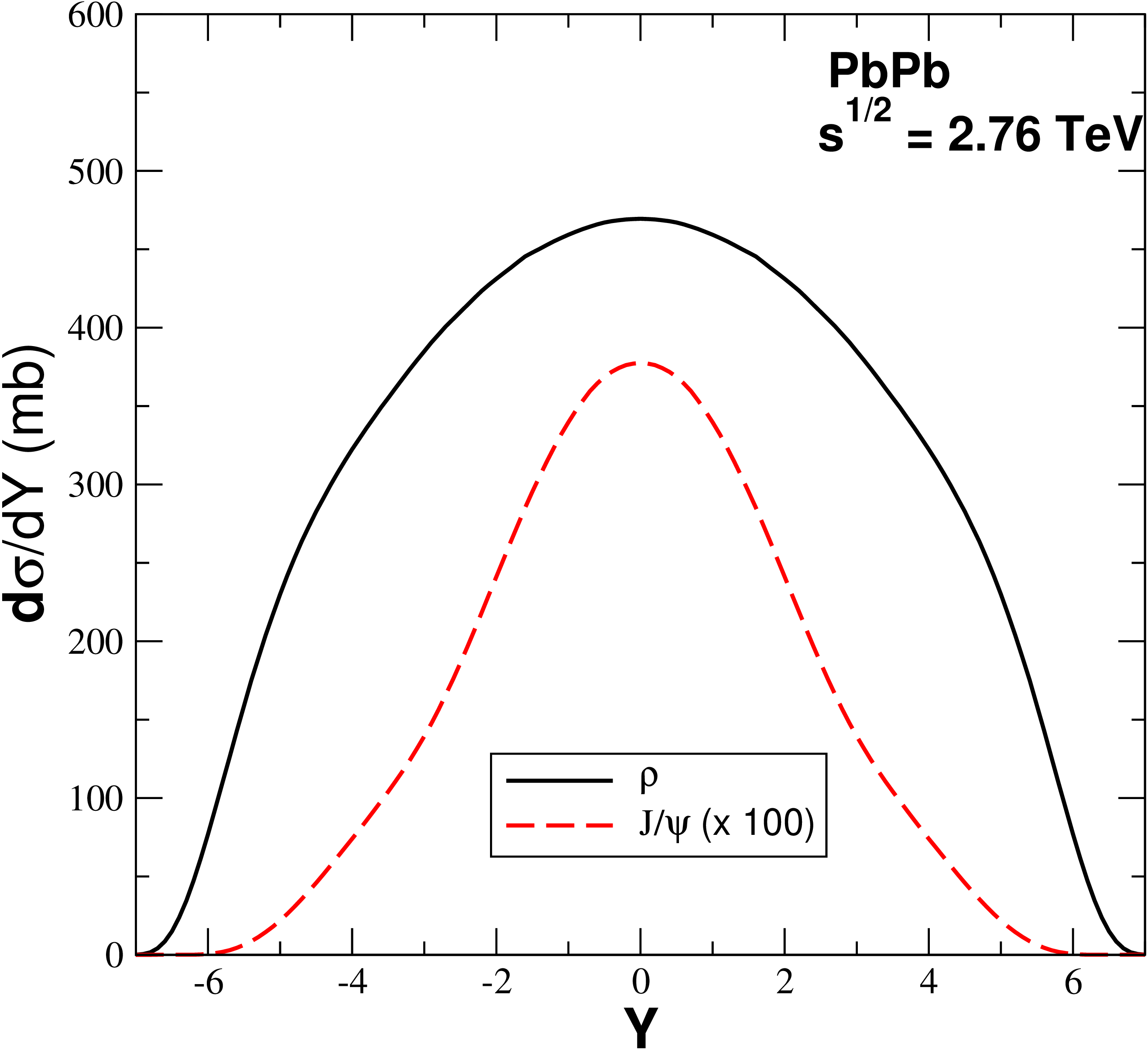} 
 \caption{(Color online) Predictions for the rapidity distribution of $\rho$ and $J/\Psi$ photoproduction in $PbPb$ collisions at LHC. }
\label{fig:3}
\end{figure}

In Fig. \ref{fig:1} we present our predictions for $pp$ collisions at $\sqrt{s} = 7$ TeV.  The results  are shown for the full rapidity range. The rapidity distribution at LHC probes a large interval of photon-proton center of mass energy since $W^2_{\gamma h}\simeq M_{V}\,\sqrt{s}\,\exp({\pm Y})$, which corresponds to very small $x\simeq M_{V}\,e^{-Y}/\sqrt{s}$. Therefore, its  experimental analysis can be useful to determine the QCD dynamics. As a reference, one has at central rapidity $\frac{d\sigma}{dy}(y=0)\simeq 6.5 $ nb (18 pb) for $J/\Psi$ ($\Upsilon$). In Table \ref{tabmesons} we present our estimates for the integrated cross sections and production rates
assuming  the  design luminosity ${\cal L}^{\mathrm{pp}}_{\mathrm{LHC}} = 10^7$ mb$^{-1}$s$^{-1}$. In comparison with previous results presented in \cite{vicmag_quarkonium} our predictions at $\sqrt{s} = 7$ TeV are almost a factor 2 (4)  smaller than those obtained for $J/\Psi$ ($\Upsilon$) production at the full LHC energy. The difference between the factors for $J/\Psi$ and $\Upsilon$ production is directly associated to the distinct energy dependence predicted by the saturation physics, which implies larger effects in the $J/\Psi$ production.

In Fig. \ref{fig:2} and Table \ref{tabmesons} we present our estimates for the $\rho$ and $J/\Psi$ production in ultraperipheral heavy ion collisions at RHIC. We assume ${\cal L}_{\mathrm{RHIC}}  = 0.4$ mb$^{-1}$s$^{-1}$.
 The results for $\rho$ production are consistent with those presented in Ref. \cite{vicmag_rho}. In the case of $J/\Psi$ production is the first time that we present the predictions using the IIM model \cite{IIM}  as input in the calculations of the dipole - nucleus cross section.
In comparison with the results presented in Ref. \cite{vicmag_mesons_per},   where the GBW model \cite{GBW}  was used as input of the calculations, our predictions are $ \approx 10 $ \% larger, which is directly associated to the difference between the description of the linear regime proposed by these two models.  In the right panel of the Fig. \ref{fig:2} we present our prediction for the  dependence on the   rapidity for the ratio between the $J/\Psi$ and $\rho$ cross sections. This observable is currently under analysis by the STAR collaboration \cite{janet}. We are considering here the case without  mutual nuclear excitation. The correction factor for meson production accompanied by mutual excitation is rather rapidity dependent and gives an overall suppression of $1/10$ in the integrated cross sections (almost the same for light and heavy mesons) \cite{Nystrand}. In the case presented here, the ratio $J/\Psi/\rho$ should have small sensitivity to those corrections and gives at central rapidity $\frac{d\sigma (\rho^0)}{dy}/\frac{d\sigma (J/\Psi)}{dy}\simeq 1.2\times 10^3$. This is consistent with the  ratio between the normalizations of the cross sections for vector meson production, which  are proportional to $(M_V^3/m_q^8)\Gamma(V\rightarrow e^+e^-)W^{4\lambda_V}$, where $\lambda_V$ is the effective energy power for each vector meson and $m_q$ is the relevant quark mass.

\begin{table}
\begin{center}
\begin{tabular} {||c|c|c|c||}
\hline
\hline
Meson & {\bf RHIC ($AuAu$)} &  {\bf LHC ($PbPb$)} & {\bf LHC ($pp$)}  \\
\hline
\hline
$\rho$ & 609.7 mb (256.0)  & 4276 mb (1796.0) &  --- \\
\hline
 $J/\Psi$ & 0.51 mb  (0.20)    &  20 mb (8.40) & 63.70 nb  (637.0) \\
\hline
$\Upsilon$ & --- & --- & 0.18 nb (1.80) \\
\hline
\hline
\end{tabular}
\end{center}
\caption{\ The integrated cross section (events rate/second) for vector meson photoproduction  in $pp$ and $AA$  collisions at RHIC and LHC energies.}
\label{tabmesons}
\end{table}

Finally, in Fig. \ref{fig:3} and Table \ref{tabmesons} we present our predictions for the rapidity distribution of $\rho$ and $J/\Psi$ photoproduction in $PbPb$ collisions in the first heavy ion run of the LHC ($\sqrt{s} = 2.76$ TeV). We assume the  design luminosity ${\cal L}^{\mathrm{PbPb}}_{\mathrm{LHC}} = 0.42$ mb$^{-1}$s$^{-1}$. In comparison with the previous results \cite{vicmag_rho,vicmag_quarkonium}, our predictions for $\rho$ production are a factor 1.4 smaller than those obtained for the full LHC energy. In the  $J/\Psi$ case, our predictions are smaller by a factor about two.  These differences are associated to the distinct contribution of the saturation effects for the $J/\Psi$ and $\rho$ production, which affects the energy dependence of the cross sections. As a reference, one has at central rapidity $\frac{d\sigma}{dy}(y=0)\simeq 470 $ mb (3.8 mb) for meson $\rho$ ($J/\Psi$).

As a summary, in this letter we updated our predictions for the vector meson production in coherent interactions at RHIC and LHC. In particular, we furnished the predictions for the center-of-mass energies of the first runs of the LHC. Moreover, we present our predictions for the dependence on the   rapidity for the ratio between the $J/\Psi$ and $\rho$ cross sections, which is currently under analysis. Our results demonstrate that the production rates are large at LHC, which implies the experimental study of this process is feasible.

\begin{acknowledgments}
 This work was  partially financed by the Brazilian funding agencies CNPq and FAPERGS. The authors thank Joakim Nystrand for useful comments and discussions. 
\end{acknowledgments}

\end{document}